\documentclass[letterpaper,twocolumn,10pt]{article}

\usepackage{usenix}
\usepackage[utf8]{inputenc}
\usepackage[american]{babel}
\usepackage{xspace}
\usepackage{amsmath}
\usepackage[makeroom]{cancel}
\usepackage{cleveref}
\usepackage{graphicx}
\usepackage{biblatex}
\usepackage{listings}
\usepackage{tikz}
\usepackage{pgfplots}
\usepackage{siunitx}
\usepackage{mdframed}
\usepackage[explicit]{titlesec}
\usepackage{subfig}

\mdfdefinestyle{sframe}{%
  skipabove=0.25\baselineskip,%
  skipbelow=0.25\baselineskip,%
  linecolor=black,%
  linewidth=.7pt,%
  leftmargin=0pt,%
  rightmargin=0pt,%
  innerleftmargin=0.5em,%
  innerrightmargin=0.5em%
}

\titlespacing*{\section}{0pt}{9pt}{4pt}
\titlespacing*{\subsection}{0pt}{7pt}{3pt}
\titlespacing{\paragraph}{%
  0em}{
  0.3\baselineskip}{
  0.5em}
\titleformat{\paragraph}[runin]{\normalfont\bfseries}{\thetitle}{1em}{#1.}

\definecolor{base03}{HTML}{002B36}
\definecolor{base02}{HTML}{073642}
\definecolor{base01}{HTML}{586E75}
\definecolor{base00}{HTML}{657B83}
\definecolor{base0}{HTML}{839496}
\definecolor{base1}{HTML}{93A1A1}
\definecolor{base2}{HTML}{EEE8D5}
\definecolor{base3}{HTML}{FDF6E3}
\definecolor{yellow}{HTML}{B58900}
\definecolor{orange}{HTML}{CB4B16}
\definecolor{red}{HTML}{DC322F}
\definecolor{magenta}{HTML}{D33682}
\definecolor{violet}{HTML}{6C71C4}
\definecolor{blue}{HTML}{268BD2}
\definecolor{cyan}{HTML}{2AA198}
\definecolor{green}{HTML}{859900}
\newcommand{\stkout}[1]{\ifmmode\text{\sout{\ensuremath{#1}}}\else\sout{#1}\fi}
\newcommand{\thesystem}{RFQuack\xspace}

\begin{document}

\date{}
\author{
  {\rm Federico Maggi}\\%
  Trend Micro Research\\%
  {\rm\ttfamily federico@maggi.cc}%
  \and
  {\rm Andrea Guglielmini}\\%
  Politecnico di Milano\\%
  {\rm\ttfamily andrea.guglielmini@mail.polimi.it}
  }

\title{\Large \bf \thesystem: A Universal Hardware-Software Toolkit \\
for Wireless Protocol (Security) Analysis and Research}

\maketitle

\begin{abstract}
Software-defined radios (SDRs) are indispensable for signal reconnaissance and physical-layer dissection, but despite we have advanced tools like Universal Radio Hacker, SDR-based approaches require substantial effort.
Contrarily, RF dongles such as the popular Yard Stick One are easy to use and guarantee a deterministic physical-layer implementation.
However, they're not very flexible, as each dongle is a static hardware system with a monolithic firmware.

We present \thesystem, an open-source tool and library firmware that combines the flexibility of a software-based approach with the determinism and performance of embedded RF frontends.
\thesystem is based on a multi-radio hardware system with swappable RF frontends, and a firmware that exposes a uniform, hardware-agnostic API.
\thesystem focuses on a structured firmware architecture that allows high- and low-level interaction with the RF frontends.
It facilitates the development of host-side scripts and firmware plug-ins, to implement efficient data-processing pipelines or interactive protocols, thanks to the multi-radio support.
\thesystem has an IPython shell and 9 firmware modules for: spectrum scanning, automatic carrier detection and bitrate estimation, headless operation with remote management, in-flight packet filtering and manipulation, MouseJack, and RollJam (as examples).

We used \thesystem to setup RF hacking contests, analyze industrial-grade devices and key fobs, on which we found and reported 11 vulnerabilities in their RF protocols.
\end{abstract}

\section{Introduction}
\label{sec:introduction}
The increased adoption of wireless communication puts security research on the front line.
Previous work has showed that both legacy~\cite{industrial_radios} and newer-generation protocols (e.g., LoRaWAN~\cite{lorawan_ioactive}) require in-depth security auditing of the RF protocols.
The impact of vulnerabilities on legacy protocols are particularly relevant, because they can affect industrial devices, which have long life spans (decades), and thus may never be patched until replacement.

After an almost-mandatory, blind replay-attack test with an SDR, the typical workflow to analyze an unknown wireless protocol begins with a quick assessment using RF dongles (embedded RF transceivers), trying to sniff messages and "play around" with them.
If the RF dongle fully supports the lower communication layers of our target protocol, then we have immediate access to the payload for further analysis.
However, there exist many low-layer implementations, each with their own peculiarities, especially if we're looking at industrial applications.
At this point we need an SDR to capture the signal and analyze it offline, with the goal of reverse engineering most of the lower communication layers.
Once we have complete knowledge of the protocol, we can start looking for flaws.
At this point it is not uncommon to bring RF dongles back in the game, because when
the protocol is fully known, it's much more reliable to use a hardware transmitter to forge messages.
Or maybe we need to quickly build a custom dongle, because we need a peculiar transceiver that supports a very uncommon modulation scheme.

Even if offline signal analysis is easier than in the past---thanks to advanced SDR tools like URH~\cite{urh}---it's still quite challenging and error-prone to develop full, dynamic, precise, reliable transceivers.
We wish we had flexible RF dongles that can be quickly reconfigured and adapted to support virtually any protocol, like with SDRs.
This is what motivated us to develop and release \thesystem\footnote{\url{https://github.com/rfquack}}, which we like to think of as "the Arduino for RF researchers."

\begin{figure}[t]
  \centering
  \includegraphics[width=0.45\columnwidth]{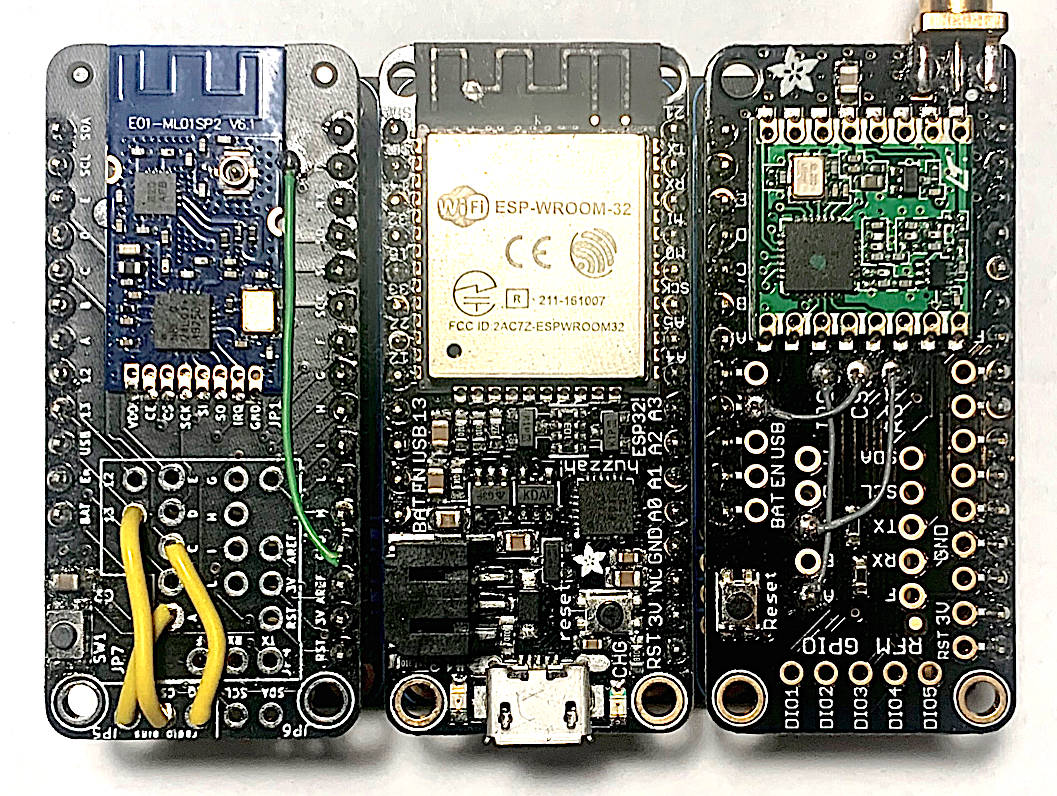}
  \includegraphics[width=0.45\columnwidth]{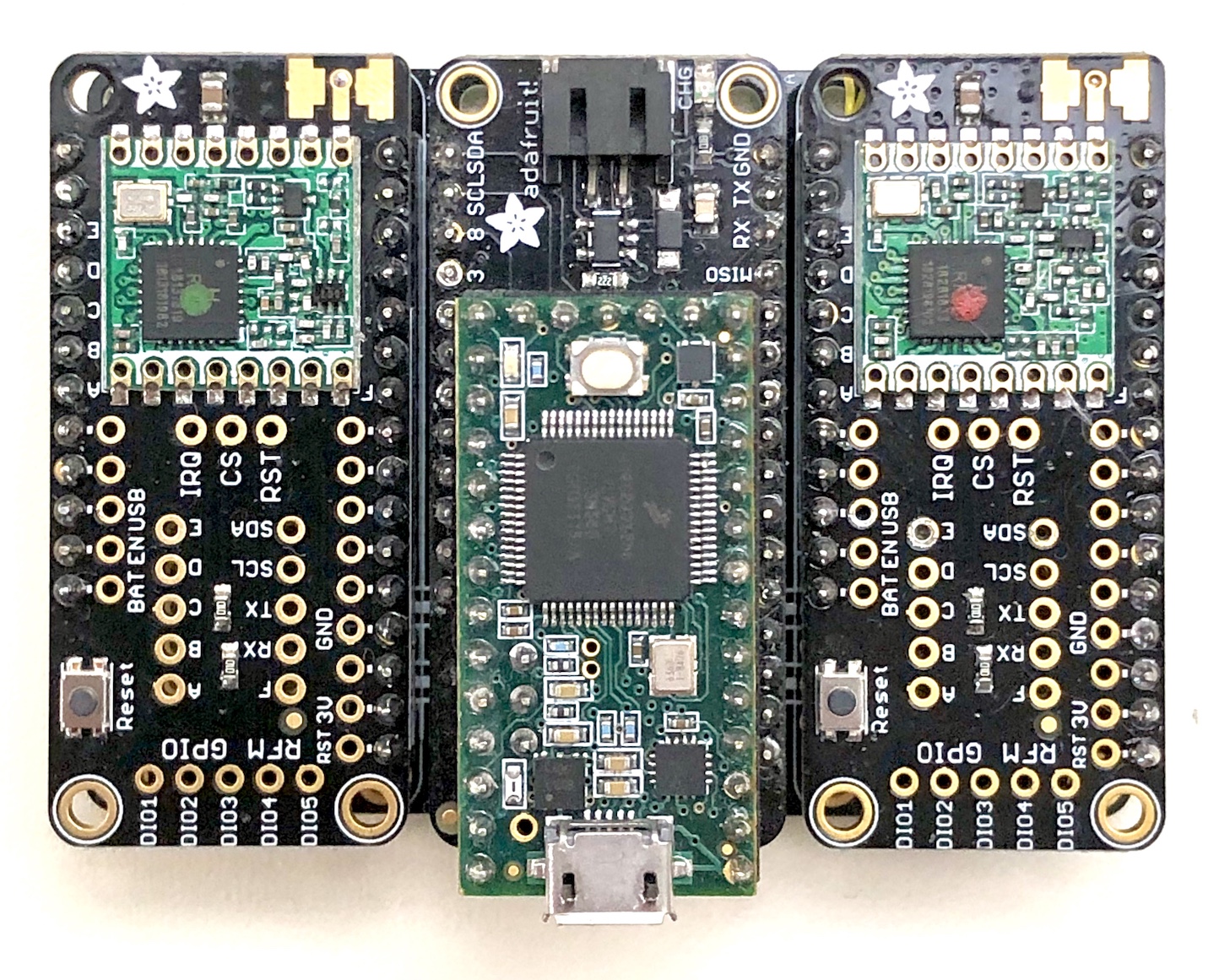}
  \caption{Two \thesystem modular dongles with main board (ESP32 and Teensy 3.2) and two RF daughterboards each (nRF24 + RF69 and two LoRa RF96, at 433 and 868 MHz).}
  \label{fig:rfq-dongles}
\end{figure}

\thesystem is a RF dongle \textit{system} with an extensible hardware and software platform that provides a solid foundation to develop custom RF dongles (see \Cref{fig:rfq-dongles}) for wireless reverse engineering.
While allowing full manual control, \thesystem includes ready-to-use RF-analysis features such as an automatic frequency and bitrate estimation that detects and clamps on transmissions in real time (in under \SI{33}{ms}, with a \SI{20}{kHz} accuracy).
It supports multiple radios simultaneously, including mixed sub-GHz and 2.4GHz ones, and virtually any embedded radio.
Also, \thesystem makes it easy to program interactive RF protocols without changing the firmware: It includes a packet filtering and modification engine that runs on the dongle and can be scripted from the host.

We verified with practical use cases that the modular firmware makes it easy to implement new functionalities.
This makes \thesystem suitable not only for security research, but also for hacking contests and trainings.
In addition to and advanced IPython shell that leverages a robust Protobuf-based RPC, we believe that its open API will ease its integration with software such as URH, to offer a hybrid RF-analysis platform based on SRD and equally-flexible RF transceivers.

\medskip\noindent In summary, we make the following contributions:
\begin{itemize}
  \item A complete and extensible open hardware and software system that makes hardware-assisted RF analysis flexible and easier to approach.
  \item We implement 9 firmware modules both to show how to use \thesystem's API and to provide the essential functionalities to RF analysis and red-team tasks.
  Among these modules we include a signal-clamping routine that automatically decodes sub-GHz signals.
  \item Through a series of case studies on real consumer and industrial RF devices, we demonstrate \thesystem's functionalities and provide a practical reference to its internals, to help future developers to extend it.
\end{itemize}

\section{State of the Art and Motivation}
\label{sec:motivation}
The essential functionality of any RF-analysis system is to correctly "translate" a bit stream into its analog representation, by modulating the carrier waveform signal (or vice versa), and let users to easily "tap" into (or control) this process.

Currently available research tools can broadly divided into SDRs and RF dongles.
These approaches have opposite benefits and drawbacks, which motivated us in creating \thesystem to bridge this gap.

\begin{table*}[t]
  \centering
  \caption{Feature comparison of SDR-based vs. RF-dongle-based approaches and tools, discussed in \Cref{sec:sdr-vs-rf}}
  \label{tab:sdr-vs-rf}

  \resizebox{\textwidth}{!}{%
  \begin{tabular}{@{}|r|c|c|c|c||c||@{}}

    \hline

   & \multicolumn{2}{c}{\textbf{SDR}} & \multicolumn{3}{|c|}{\textbf{RF Dongles}} \\

    \hline

  & \textbf{GNU Radio} & \textbf{URH} & \textbf{Single Radio} & \textbf{Multi Radio} & \textbf{\thesystem} \\
    \hline

  \textbf{Spectrum Coverage} & \multicolumn{2}{c|}{$10^{3}$ to $10^{6}$Hz, continuous} & \multicolumn{2}{c||}{Only specific bands} & Arbitrary bands, discrete \\
    \hline
  \textbf{RF Parameters} & \multicolumn{2}{c|}{Any (custom physical layer)} & \multicolumn{2}{c||}{Hardware bound} & All (modular) \\
    \hline
  \textbf{Host Interfaces} & \multicolumn{2}{c|}{Wired (USB, Ethernet)} & USB, BLE & USB, BLE, WiFi, Cellular & Any (modular) \\
    \hline
  \textbf{API Uniformity} & \multicolumn{2}{c|}{High (established project)}  & \multicolumn{2}{c||}{Very low} & High \\
    \hline

  \textbf{Extensibility} & Very high & High & \multicolumn{2}{|c||}{Very low (SW)} & High (HW, SW) \\
    \hline
  \textbf{Main Target Layers} & Physical & Logic & Logic & Hybrid & Hybrid \\
    \hline
  \textbf{Focus} & Research & Research, RE, Fuzzing & Red-teaming & Red-teaming & Research, RE, Fuzzing \\
    \hline
  \textbf{Development Effort} & High (DSP) & Medium (UI, Scripting) & Low (Scripting) & Low (UI, Scripting)  & Medium (Scripting, C) \\
    \hline
  \textbf{Transceiver Performance} & \multicolumn{2}{c|}{Medium (Latency)} & Very high (IC) & High (SPI) & High (SPI) \\
    \hline
  \textbf{Hardware cost} & \multicolumn{2}{c}{\$20–15,000} & \multicolumn{3}{|c|}{\$10-200} \\
    \hline

  \end{tabular}%
  }
  \end{table*}

\subsection{Software-defined Radios (SDRs)}
\label{sec:sdr}
To receive or transmit data with SDRs, we must implement most of the wireless protocol, down to the physical layer.

\paragraph{Hardware vs. Emulation} Traditional radio systems have hardware parts such as filters, converters, modulators, and demodulators, connected together to process any signal received by the antenna, and translate it into a different form (e.g., audio)---and vice versa.
In SDRs, such components are emulated in software, so they can implement any (wireless) communication stack without changing hardware.

\paragraph{High-end SDRs} While basic SDRs are essentially analog-to-digital converters (like the popular RTL2832U DVB-T receiver), professional SDRs integrate advanced digital signal processors (DSPs), field programmable gate arrays (FPGAs), and other dedicated, programmable hardware, that run the intensive tasks.
The pure software part of such SDRs just configures the board and emulates the remainder components not available on the board.

\paragraph{Signal Sampling} Once tuned at a given frequency and bandwidth, an SDR board captures sample values of any received electric current, which is an analog, alternating signal (also known as \textit{waveform}).
These samples form a time series of complex values---often known as "I/Q data," where I and Q are, respectively, the real and imaginary parts.
SDRs capture (or generate) millions of samples per second (sps), producing a high-resolution digital representation of an analog signal waveform.

\paragraph{Post-processing} The received I/Q samples are transferred to the host via Ethernet or USB links for processing, with tools such as GNU Radio~\cite{gnuradio} or URH, which facilitate demodulating, interpreting, and decoding any (digital) data carried by the waveform.
Regardless of the methodology, the outcome is a bit stream, which can be parsed for higher-level protocol dissection.

\subsection{Embedded RF Dongles}
\label{sec:rf-dongles}
To receive or transmit data with RF dongles, one must obtain or create a dongle that matches (i.e., implements) the physical layer of the target signal or device.

\paragraph{True Radios} RF dongles are small embedded systems with a host interface (typically serial, via USB or BLE), a micro-controller unit (MCU), and a digital transceiver module.
The first and most-popular hacker-friendly RF dongle is the Yard Stick One~\cite{ossmann2016rapid}, based on the Texas Instrument CC1111 chip, which integrates, within a single package, an MCU and the CC1101 sub-GHz radio module.
The MCU runs a firmware that let host client software change RF parameters and transmit or receive data, by interacting with the registers on the MCU and the RF module.
Unlike SDRs, RF dongles implement a protocol's physical layer \textit{in hardware}, which processes the analog signals received by the antenna into bit streams---and vice versa---exposing a digital communication interface to the host.

\paragraph{Decoding Capabilities} The capabilities of an RF dongle is bound to its transceiver, which cannot be changed since its soldered or integrated.
For example, the Yard Stick One, makes it straightforward to decode data modulated with ASK or FSK, but PSK is simply not supported by the CC1101.
Unlike with SDRs, the analyst must know the RF parameters (i.e., modulation scheme, carrier frequency, and bitrate) \textit{a priori}, as the (de)modulation happens in hardware.

\paragraph{Post-processing} The firmware running on an RF dongle "talks" with a companion client tool on the host, typically a mobile application or command-line interface (CLI) that makes the dongle actually usable.
RfCat~\cite{rfcat} is the first and most-popular CLI for CC1111-based RF dongles such as the Yard Stick One or the PandwaRF~\cite{pandwarf}.
Without going into further details, there exist similar solutions for the 2.4GHz bands: The nRF24 is a popular hacker-friendly radio transceiver, which has inspired the creation of various RF dongles, firmware, and utilities (e.g., nRF Research Firmware~\cite{nrf_research_firmware}).

\subsection{Motivation: The Best of Both Worlds}
\label{sec:sdr-vs-rf}
The diverse characteristics of SDRs vs. RF dongles summarized in \Cref{tab:sdr-vs-rf} have motivated researchers and hardware developers to create feature-rich boards with multiple transceivers that cover the most common bands.
For instance, the HackCube integrates an SDR receiver, WiFi, Bluetooth, NFC, and a 2.4GHz and sub-GHz transceiver.
While tools such as the HackCube have merit, they are still monolithic (for miniaturization reasons).
Without an open, flexible design, it's hard for the community to maintain and contribute to the development of such devices: Adding hardware or software modules to devices like the HackCube is possible by only patching it, as it's not designed for extensibility.

The \fbox{framed text} in this section indicates an observation or design principle beneath \thesystem.

\paragraph{Spectrum Coverage and RF Parameters}
The main characteristic of SDRs is their wide spectrum coverage (or bandwidth).
Even a \$20 SDR can tune at any frequency between 300MHz and 1.5GHz.
By design, RF dongles can only tune to a very limited, discrete set of bands (e.g., 125kHz, 315MHz, 433MHz, 2.4GHz) because their RF frontends have a basic frequency synthesizer.

Since the physical layer is implemented in hardware, RF dongles can only support a finite set of modulation schemes, synchronization words, packet formats, and CRC algorithms.
\begin{mdframed}[style=sframe]
Adding new frontends for RF dongles to support new bands and physical layers should be as easy as connecting the right hardware module, like in advanced SDR systems (e.g., Ettus USRP), which come with a variety of daugtherboards.%
\end{mdframed}

\paragraph{Host Interfaces}
SDRs need high-bandwidth interfaces such as USB or Ethernet.
The most popular RF dongle (Yard Stick One) is based on USB, while the more recent PandwaRF, WHID Elite, or HackCube, have BLE, WiFi or cellular interfaces, mostly because they are conceived with a red-teaming purpose (e.g., physical security), so they can launch attacks remotely, in a headless fashion.

\begin{mdframed}[style=sframe]
If RF the many dongles had a uniform connectivity layer on top of such interfaces---instead of custom, undocumented protocols---researchers would be empowered to develop with these platforms and benefit from their diverse features.%
\end{mdframed}

\paragraph{API Uniformity}
GNU Radio offers a uniform, documented API.
While performance-critical signal-processing blocks can be written in C++, GNU Radio allows to develop radio applications in Python, and compose them with a GUI.
URH exposes a plugin API to extend its signal-processing, decoding, and fuzzing capabilities.

\begin{mdframed}[style=sframe]
The only effort to provide a uniform API in the RF dongles world is PandwaRF's Android API and SDK.
However, being the PandwaRF a commercial, single-radio product, its firmware is not open source and the hardware capabilities are bound to the fixed RF frontend.
Similar efforts should be embraced by open source RF dongle systems.%
\end{mdframed}

\paragraph{Extensibility and Development Effort}
Research tools like URH make RF reverse engineering accessible, although implementing complex protocols is challenging: Despite GNU Radio offers a high-level experience with fine-grained control over the processing pipeline, RF protocol design calls for in-depth DSP knowledge.

While it is easy to extend and port software across different SDR hardware thanks to the uniform API exposed by toolkits like GNU Radio or URH, the RF world is very different: Each system has its own custom firmware.
As a result, even when the firmware is open sourced, it is hard to extend unless by patching it.

\begin{mdframed}[style=sframe]
Extending the firmware of an RF dongle should be facilitated by a consistent API, which allows the developers to focus on adding new functionalities, rather than reinventing custom abstraction layers.
\end{mdframed}

\begin{figure*}[t]
  \centering
  \includegraphics[width=0.9\textwidth]{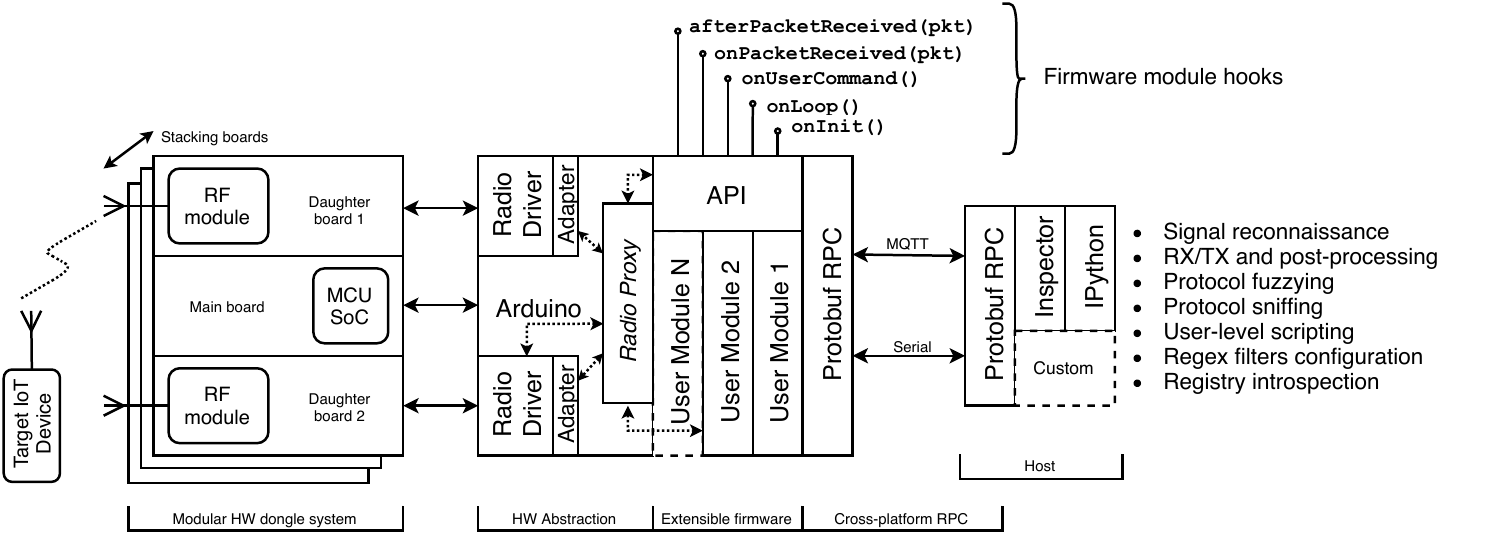}
  \caption{The high-level architecture of \thesystem presented in \Cref{sec:architecture} with details in \Cref{sec:details}.}
  \label{fig:architecture}
\end{figure*}

\paragraph{Main Target Layers}
SDRs allow to work at any protocol layer.
Although tools like GNU Radio are meant primarily to work at the physical layer, they have support to parse the logic of digital protocols (e.g., packet format).
Instead, URH is focused on supporting the reverse engineering of the \textit{logic} layer.

RF dongles only allow to interact with the logic layer of protocols.
Depending on the RF frontend, some RF dongles may allow partial access to lower layers.
For instance, some RF dongles allow to consume RSSI\footnote{The receiver signal strength indicator is the measured power of a received signal at a given carrier frequency.} samples fast enough to estimate the spectrum density.
In \Cref{sec:auto-tuning} we show how we use these samples to implement the auto-tuning feature of \thesystem.
The HackCube even includes a simple SDR receiver, although creating custom receivers is still constrained by the MCU computational power.

\paragraph{Focus}
Both SDRs and RF dongles can be used for research purposes, but there are some small differences.
Out of the box, GNU Radio is less fit for offensive tasks, although it does not prevent to develop custom, even very advanced, physical-layer fuzzing routines.
URH is conceived with a clear RE orientation: Besides the UI that assists protocol parsing, URH has ready-to-use functions to create fuzzing templates and them directly against the targets.

RF dongles are optimized for offensive and red-team activities.
For example, the PandwaRF has rolling-code-cracking routines, WHID Elite~\cite{whidelite} comes with a modem (to simulate remote attackers) and a USB port (for HID exploits).

\begin{mdframed}[style=sframe]
To draw an analogy, \thesystem is more inspired by URH than by the available RF dongles: It facilitates the extension of its firmware to support custom protocol parsing, complex tasks such as fuzzing or interactive logics, and to test for known vulnerabilities (e.g., RollJam, MouseJack).%
\end{mdframed}

\paragraph{Development Effort}
It is not only hard to develop physical-layer protocols (e.g., with GNU Radio), but it requires in-depth knowledge in analog and digital signal processing, information theory, and cryptography.
Frameworks like GNU Radio are very sophisticated and focused on technical advances rather than on \textit{user} documentation and usability, which make them not easy to approach.
URH has made a step forward, by providing a modern UI that works out of the box, a plug-in API, and scripting capabilities.
RF dongles make "quick" tasks very easy to approach, as they offer a simple UI and Python scripting, but it is not immediate to develop firmware extensions, and nearly impossible to change their hardware (most RF modules are integrated or embedded on the board).

\paragraph{Transceiver Performance}
Single-radio RF dongles can embed the radio frontend and the MCU in a single, integrated SoC package, which minimizes latency on the bus.
While SDRs can be optimized with FPGA or hardware DSPs, a non-negligible portion of the protocol consists of native or interpreted (Python) code running on the host computer, on top of a general-purpose OS, which cannot comply with the determinism and real-time constrains required to implement accurate, efficient, and reliable transceivers.

\begin{mdframed}[style=sframe]
While \thesystem's design imposes some latency due to the detachable RF frontends, the physical layer is implemented in hardware, which guarantees far more precision and reliability than any SDR implementation.
Some protocols---especially in the 2.4GHz bands and above---simply cannot be conveniently developed with an SDR, if not with very expensive hardware and FPGA offloading.
\end{mdframed}

\section{\thesystem High-level Overview}
\label{sec:overview}
Both hardware and software of \thesystem are fully modular.
This ensures that we can always leave convenient "tapping" points for users to customize,
while relying on a uniform interface to multiple radio transceivers.

In addition to using the built-in IPython shell, we make it easy for users to build firmware modules to implement signal-reconnaissance tasks, transmit or receive (post-processed) data, discard packets not matching filters (e.g., selective protocol sniffing), as well as interacting at the low level with the MCU and the transceiver (e.g., read and write registers).

\subsection{The Journey of a Wireless Transmission}
\label{sec:architecture}
Reading \Cref{fig:architecture} from left to right, (1) when a target device (e.g., an IoT embedded system) transmits packets, the signal is received by one or more of the daughterboard radios.
An interrupt request from daughterboard (2) invokes \lstinline{onPacketReceived(pkt)} of all loaded module, in a user-configurable order.
For example, if enabled, (3) the packet-filtering module ignores packets not matching a set of user-provided regular-expression patterns, and enqueues the accepted ones for further processing.
If enabled, (4) the packet-manipulation module pulls packets from such queue, (5) pipes them through a set of user-provided modification rules, and (6) enqueues them again.
The repeater module, if enabled, (7) consumes any enqueued packet and (8) pushes them to another queue.
Depending on the settings, (9) packets can be serialized and proceed towards the host.
Alternatively, the user may set \thesystem to (11) forward the (modified) packets to the repeater module, which immediately transmits them.

\subsection{Modular Hardware Design}
\label{sec:hw-modularity}
It is possible to swap \thesystem daughterboards because they're connected with simple pin headers.
Multiple boards, with up to 4 RF daughterboards each, can be stacked using a simple pin header.
Each daughterboard needs a serial peripheral interface (SPI) bus, plus some optional GPIO pins.
The 3 SPI bus lines (SI, SO, CLK) are shared across all daughterboards (SPI slaves).
To allow the MCU (SPI master), to selectively talk to a specific transceiver, each RF daughterboard independently routes 1 slave-select (SS) line to the main board, plus 1 or more GPIO lines (e.g., for interrupts).
The MUC triggers the SS to notify a specific transceiver that there is data on the SPI bus.
The transceivers can start a SPI transaction by triggering an interrupt on a GPIO.

\subsection{Modular Firmware Design}
\label{sec:sw-modularity}
As shown \Cref{fig:architecture}, the module API exposes 5 methods, which allow to implement user modules that can tap into the data-processing pipeline: \lstinline{onInit()}, when the module is first loaded; \lstinline{onLoop()}, at each MCU cycle; \lstinline{onUserCommand()}, when a valid command is received via the RPC; \lstinline{on/afterPacketReceived(pkt)}, called upon and after a packet is received from the transceiver.
Each functionality of \thesystem is implemented as a module.
For example, the radio-management module receives and dispatches users commands (\lstinline{onUserCommand()}) such as set/get registers and set/get RF parameters.
The packet-filtering module, instead, is triggered on the \lstinline{onPacketReceived(pkt)} event, and calls \lstinline{onLoop()} to flush any received or manipulated packet as soon as possible.
In \Cref{sec:rolljam} and \ref{sec:mousejack} we showcase the module API by implementing RollJam and MouseJack, two popular RF attacks.

\paragraph{Transceiver-agnostic, Uniform API}
Like with GNU Radio sink blocks \thesystem's abstracts different transceivers via the same API, which provides functions to change the radio mode (RX, TX, idle, jam, promiscuous), RF parameters (carrier, bitrate, deviation, bandwidth), packet format, output power, to get-set registers, and to send-receive binary data.
The list of exposed functions is in \Cref{sec:uniform-api}.
Optionally, it exposes transceiver-specific functions (e.g., the CC1120 has 16 bit extended registers).

\paragraph{Connectivity and Headless Operation}
To meet red-teaming operational requirements---although more compact RF dongles exist---\thesystem is designed with an open connectivity model in mind.
\thesystem has built-in support for serial (USB) and MQTT transport (for remote, distributed setups), and the client side supports deserialization and automatic type inference.

\subsection{Automatic Signal Clamping}
\label{sec:auto-tuning}
In addition to the precise carrier frequency of the target communication, we must know its data rate (or bitrate, bits/sec) to correctly interpret a demodulated signal into bits.
Usually, when short of luck at inspecting FCC (leaked) documents, we employ an SDR and a spectrum analyzer to measure the power value at each frequency, and visually identify the highest peak.
Similarly, if the bitrate is unknown, the only option available when using RF dongles is to set the bitrate to the highest value, set the radio in promiscuous mode (i.e., do not filter incoming packets based on preamble length or sync-word), and post-process any captured data to downsample it to the correct bitrate.
This is not ideal, because it does not allow the user to fully leverage on-board filters and other functionalities of the firmware.

\thesystem has a \textbf{frequency-finder module} that listens on a range and, as soon as it detects a new transmission, tunes to that frequency and triggers the \textbf{bitrate-estimator module}.
For example, if a car's key fob is pressed within range, the modules infers carrier frequency and bitrate, and present the user with the payload.
This happens in real time, entirely on the dongle, without interfering with other tasks.
It is important to clamp on the signal before its preamble ends, so that \thesystem will be able to synchronize and decode the transmitted payload.
Although \thesystem supports multiple radios, these modules uses a single transceiver, leaving any spare transceiver available to work on other frequencies.

While the PandwaRF supports similar functionalities, we're the first to implement it as a module and, most importantly, release it as open source code.
\section{Implementation Details}
\label{sec:details}
\thesystem modular hardware is based on Adafruit's Feather system, as it comes with stacking boards with up to 4 slots each.
We tested early versions of \thesystem on the ESP32 and ESP8266 SoCs (MCU + WiFi), the RFM69 and CC1120 RF modules (sub-GHz bands), and the SIM800 cellular modem.
We provide the open-hardware schematics for a CC1120 adapter and a compact nRF24 FeatherWing.
The most recent version supports the CC1101 and nRF24 RF modules\footnote{We're currently porting from RadioLib the wrapping code to support the RFM2x, RFM69, and RFM9x LoRa modules.}, working simultaneously, with up to 5 radios (a shared SPI bus plus 2 GPIOs are required for each radio), and we tested it with the Teensy 3.2.
Thanks to the Arduino and similar abstraction frameworks, \thesystem could run on virtually any of the 800 boards supported by PlatformIO\footnote{\url{https://platformio.org/boards}}.

\subsection{Loop, Dataflow and Queues}
\label{sec:dataflow}
\thesystem's firmware runs a high- and a low-priority loop.
Any new packet goes through all the high priority tasks and then, via a decoupling queue, through the low priority ones.
Each module may implement different---high or low priority---hooks to interact with the underlying framework.


For better compatibility, we do not to leverage hardware parallelism, since it is supported only by few boards (e.g., ESP32 has two cores, while ESP8266 is single core).
However, if a user wants to trade off compatibility for speed, they can just change \thesystem's main \lstinline{loop()} to assign the high- and low-priority loops to two distinct cores.

\subsection{Multi-radio Abstraction Layer}
\label{sec:rf-abstraction}
\thesystem abstracts each RF frontend and exposes a \textit{proxy} API.
Instead of writing our own native drivers for each RF frontend, we use a decoupling driver \textit{adapter} that wraps the native driver.
We treat native drivers as external dependencies (we mainly use RadioLib~\cite{RadioLib}), which can evolve independently from \thesystem.
As \thesystem supports multiple radios simultaneously, the proxy forwards each request to the correct driver.
For example, \texttt{setModulation(OOK, RadioB)} sets the modulation of the second frontend to OOK.
Since each RF frontend may have unique features, not covered by the proxy API, we still allow direct access to the native driver.

\subsection{Carrier Frequency Detection}
\label{sec:peak-detection}
Many transceivers provide the RSSI value, an estimate of the power received, in decibel.
Samples of the RSSI can be used to draw a low-resolution spectrum (e.g., the PandwaRF Android mobile application provides that) and detect peaks, in order to spot transmitting devices nearby.
A naïve approach would be to simply loop through all the frequencies in a target range and collect RSSI samples.
We measured that it takes no less than $1ms$ for the CC1101 to tune and provide a reliable RSSI sample, which means at least $5s$ to scan the 432--437MHz range with a kHz resolution.

\paragraph{Scan by Region}
Instead of tuning to all the frequencies, we use a scan-by-region approach, leveraging the programmable receiver bandwidth filter available on any modern transceiver, which let us narrow the range of frequencies that influence the RSSI.

We set the filter bandwidth to the maximum value, $B_{max}$, take RSSI samples in the middle of each of the $N$ regions, and identify the most active one.

\begin{figure}[t]
  \centering
  \includegraphics[width=\columnwidth]{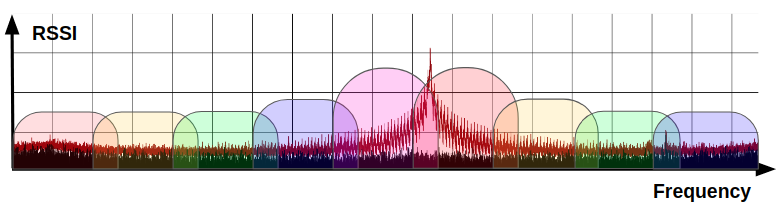}
  \caption{Carrier frequency peak detection (\Cref{sec:peak-detection}).}
  \label{fig:peak-detection}
\end{figure}

We tune to the center, $f_i$, of each region, starting from a given offset, $f_{o}$, is then:
\begin{displaymath}
f_{i} = f_{o} + i \cdot \frac{(1 - c)}{2} \cdot B_{max} \quad i \in \{0, 1, 2, \dots, N\},
\end{displaymath}
where $c \in (0,1)$ is the region-overlap ratio.
As shown in \Cref{fig:peak-detection}, the regions must overlap to avoid corner cases due to the natural attenuation of the non-ideal bandpass filters.
A guiding criterion to set the value of $c$ is that it should be proportional to the filter slope, which can be obtained from the filter response reported on the transceiver datasheet.
We obtained reliable results by setting $c = 0.25$ on the CC1101.

\paragraph{Trichotomic Search}
We run a fine-grained search within the most active region to find the peak frequency, by halving the receiver's bandwidth at each iteration.
To avoid ties while keeping efficiency, after halving the bandwidth, we split the resulting search domain in 3 sub-regions.
This narrows down the theoretical search time---within a 5MHz range---to $21 ms$, while a linear search of each of the 87 regions\footnote{The widest receiver passband filter on the CC1101 is 812kHz, which means 9 regions for the first pass and 87 iterations with progressively narrower filters, down to 58kHz.} would take $87ms$.

\paragraph{Tuning Time Optimization}
To limit the time $t_{tune}$ required by the transceiver to tune and provide stable RSSI readings, we pre-compute and cache the calibration registry values.

Most transceivers include self-calibration routines, which must be run before tuning to a frequency (or channel).
For instance, the CC1101 can hop to a frequency in $t_{hop} = 75 \mu s$ and calibrate in $t_{cal} = 712 \mu s$.
The radio driver also introduces a latency when sending the calculated registry values on the SPI bus.
We instrumented the firmware, tuned the radio 100 times on different frequencies and measured $t_{driver} = 320 \pm 20 \mu s$.

Since the range of frequencies is known in advance, we precompute and cache the calibration registry values for each frequency.
Overall, $t_{tune} = t_{hop} + \cancel{t_{cal}} + t_{driver} + t_{RSSI} = 75 + \cancel{712} + 320 + 600 ms \leq 1ms$, where $t_{RSSI}$ is the minimum time for the radio to provide a stable RSSI value.

\subsection{Automatic Bitrate Estimation}
\label{sec:bitrate-estimation}
By default, \thesystem sets the receiver to its highest supported bitrate, thus oversampling any incoming transmission.
The Nyquist–Shannon sampling theorem implies that an analog signal must be sampled at least twice its bitrate to reconstruct the rectangular wave\footnote{This limits \textit{any} bitrate-estimation technique to half of the receiver's maximum bitrate.}.
Any digital signal (especially in the sub-GHz bands) always starts with a preamble (i.e., a sequence of alternating 0s and 1s) to "wake up" the receiver, which will use it to synchronize to the incoming packet.
At this point, as soon as \thesystem has received enough oversampled alternating symbols with repetitions (e.g., four 1s followed by four 0s, repeated 3 times), it uses this array of symbols to estimate the true bitrate.
The estimated bitrate, $\hat r$, is such that the received array (oversampled at $r_o$) has only one repetition of each symbol.
We approximate this with the (weighted) average of the number of consecutive 1s (or 0s) found in the preamble $P$:
\begin{equation}
  \hat{r} = \frac{r_{o} \sum_i w_i}{\sum_i |p_i| \cdot w_i}
  \label{eq:brate}
\end{equation}
Where $|p_i|$ is the number of consecutive 1s within the preamble $P$ and $w_i$ is the number of occurrences of $p_i$ in $P$.
For example, $P = \overline{1}|\underline{00}|\overline{111}|\underline{000}|\overline{11}|\underline{000}|\overline{111}$ counts as $1 \cdot 1 + 3 \cdot 2 + 2 \cdot 1$ divided by the sum of weights, $1 + 2 + 1$.
Since the preamble does not carry any information, we accept to lose it during this operation, with the advantage of not requiring to store the packet for downsampling: We adjust the correct bitrate \textit{before} the preamble ends, so any subsequent incoming data is captured at the correct bitrate.

For simplicity, we assumed a 2-symbols OOK scheme: \Cref{sec:discussion} explains how to extend this to other modulation schemes.
The 2.4GHz is very different and this functionality is not really needed: The nRF24 transceiver, and most of the 2.4GHz transceivers, only support 2--3, very high, bitrate values.

\paragraph{Runtime Estimation}
The only operation run on the MCU is a loop over a fixed-length bit stream, $P$, to count the repetitions in $p_i$. 
The upper bound is determined by the chosen oversampling bitrate, $r_{o}$, and minimum expected preamble length, after which an estimation is available.

For example, to capture signals up to 15 kbps, we set $r_{o} = 30kbps$ and stop after 32 bytes (half the size of the CC1101’s FIFO), which translates into roughly $8.53ms = \frac{1}{30kbps} \cdot 32 \cdot 8 bits$.

\begin{table}[t]
  \caption{Key fobs used in our first case study.}
  \label{tab:key fobs}

  \centering
  \resizebox{\columnwidth}{!}{%
  \begin{tabular}{|l|l|l|l|l|l|} 
    \hline
		\textbf{Car Model} & \textbf{Key fob Model} & \textbf{Freq.} (MHz) &
		\textbf{Bitrate} (kbps) & \textbf{Transm. len.} (ms) \\
		\hline
		Fiat               & RX2TRF198             & 433.92         & 9.6      & 295              \\ 
		\hline
		Nissan             & TWB1G766              & 433.92         & 5.0      & 370              \\ 
		\hline
    Audi               & HELLA FS12A70         & 434.42         & 3.4     & 183              \\
    \hline
  \end{tabular}}
\end{table}

\section{Real-World Case Studies}
\label{sec:cases}
The goal of this section is to showcase how \thesystem applies to various scenarios, from fuzzing unknown protocols to creating custom modules that implement (known) exploits.
We have presented some these case studies as PoCs of \thesystem at the CanSecWest conference and at the Armory of Hack In The Box.

\subsection{Sniffing Key Fobs and Opening a Car}
\label{sec:key fob}
We started with cars key fobs, since they are widely available RF transmitters, they operate on the ISM bands, they are interesting targets (since they protect valuable assets from theft and robbery), and transmit a rolling code, so the payload content changes while the structure is fixed.
We used the 3 key fobs listed in \Cref{tab:key fobs}.
Also, we used a BladeRF to capture the signals emitted by these transmitters and used them as a baseline to validate the auto-clamping modules (see \Cref{sec:experiments}).

Using the IPython shell connected via serial to a \thesystem dongle, we configured the system in promiscuous mode.
Despite the RF noise, and even if we didn't know the sync word in advance, we quickly setup the packet-filtering module to discard any packet without a valid preamble.
Using the console it was easy to transmit some packets back to the receiver, which were discarded because of the rolling-code mechanism\footnote{Video of this demo: \url{https://youtu.be/jLI-oby2mu0}}.
It was during this experiment that we discovered that the receiver of one the cars was vulnerable to replay attack\footnote{We're reaching out to the car manufacturer for responsible disclosure, because the car model and brand is current and very popular.}: We were able to deterministically open the doors despite the code was rolling at any transmission.

\subsection{Vulnerable Industrial Devices}
\label{sec:industrial-radio}
We analyzed the protocol of 5 sub-GHz industrial radio devices used to automate and control manufacturing and logistic processes\footnote{We cannot explicitly mention the vendor names without indirectly revealing our names, because we are the only research group that looked at those devices and there was quite extensive media coverage. In case of acceptance we will include all the identifying details.}.

In addition to sniffing packets for analysis, we used \thesystem to implement a loop to find the correct rolling code, given a rolling code we knew from another, same-vendor device.
\begin{lstlisting}[language={Python},basicstyle=\ttfamily\scriptsize]
  q.packet_filter.add(pattern="^aaaa", negate=False)
  q.radioA.rx();  q.radioB.tx()

  for pkt in q.data:
    # we knew rcode0 from another, same-vendor device
    pkt[2:4] ^= rcode0  # XOR with known rolling code
    q.radioB.send(pkt)  # transmit
\end{lstlisting}
Once we heard the receiver accepting our command, we knew that we had used the correct rolling code.
This allowed us to exhaustively build a table of valid rolling codes.
Knowing valid rolling codes and the packet structure, we configured the filter, manipulator, and repeater modules of \thesystem to wait for one valid transmission, change some specific bytes in the packet, and transmit the modified packet.
\begin{lstlisting}[language={Python},basicstyle=\ttfamily\scriptsize]
  q.packet_filter.add(pattern="^aaaa", negate=False)
  q.radioA.reset_packet_mods()  # reset module

  # XOR byte 7 with 0x04, byte 10 with 0x08, etc.
  q.radioA.add_packet_mod(i=7,  val=0x04, op=XOR)
  q.radioA.add_packet_mod(i=10, val=0x08, op=XOR)
  q.radioA.add_packet_mod(i=12, val=0x04 + 0x08, op=XOR)
\end{lstlisting}
As a result, the vulnerable target receiver executed a command of our choice.
Moreover, since the routine was reactively executing the same loop in a headless and continuous fashion, we were able to keep the target in to a persistent denial-of-service state, by repeatedly sending "shut down" commands.
Note that the packet-modification routine runs on the MCU, and could be used to implement multi-radio scenarios like the one just presented.

\subsection{Sniffing 2.4GHz Protocols}
\label{sec:24ghz-protocols}
The two most challenging aspects of 2.4GHz protocols are that the spectrum is very crowded (e.g., cellular, WiFi, Bluetooth) and protocols can use frequency hopping.
This makes a pure SDR approach quite impractical.
We conducted this experiment outside a Faraday cage, in challenging conditions: In addition to cellular traffic, there was a WiFi AP and a Bluetooth smartwatch.
Knowing only the bitrate from the public FCC database, we were able to identify the exact frequency of a Microsoft wireless mouse and narrow down its address in seconds, with only 6 minutes of manual work on the command line (see \Cref{sec:24ghz-analysis}).

By generating several packets (i.e., by moving the mouse excessively nearby the receiver) we isolated its sync word (and confirmed it from FCC database)\footnote{Video of this demo: \url{https://youtu.be/c4OSh3jQNsY}}.
Then, by looping once again through all the frequencies, we isolated only the traffic coming from that mouse,
given that we knew the sync word.
The same experiment would have required hours of SDR development.

\subsection{MouseJack Attack Implementation}
\label{sec:mousejack}
MouseJack is a well-known attack against non-Bluetooth 2.4GHz HID devices such as mice and keyboards, which has recently enjoyed quite some attention~\cite{mousejack}, because many vulnerable, unpatched devices are still sold an used.
We have taken this popular attack as a representative example to show how to create a custom user module that implements the exploit.
The disarmed source code (i.e., without payload) is available at \thesystem's repository, so we hereby focus on how the implementation leveraged \thesystem's module API.

The MouseJack module overrides the \lstinline{onUserCommand()} to react on "start/stop" commands sent over serial or MQTT, to receive the attack payload (e.g., HID commands) from the user.
The "start" command puts the radio in promiscuous mode (using the built-in \lstinline{setPromiscuousMode()} method) and starts receiving (\lstinline{setMode(RX)} and \lstinline{setFrequency(2400)}).

The MouseJack module overrides the \lstinline{onPacketReceived(pkt)} function, to check if the scanning loop has captured a valid payload, by checking that the packets contain a valid sync-word and pass the CRC.
Based on the received valid packets, the module continues and fingerprints the transmitting victim device (e.g., Microsoft, Logitech).
The resulting module has under 350 lines of code, while the reference standalone implementation has over 470 lines, plus libraries.

\subsection{RollJam Attack Implementation}
\label{sec:rolljam}
In the sub-GHz range, RollJam is a very popular rolling-code-cracking attack.
The inventor~\cite{RollJam} used a custom-made device with two transceiver: One is to jam the legitimate receiver (i.e., on the car), and the second to listen for packets from the key fob, to capture rolling codes---which will never be received by the legitimate, jammed receiver.
Last, RollJam stops the jamming loop and transmits the captured still-fresh rolling code, which will be accepted by the receiver that couldn't see such code while blinded.
We have taken this popular attack as a second, representative example to show how to develop custom modules by leveraging \thesystem's firmware API to multiple radios.
The source code is available at \thesystem's repository, so we hereby focus on how the implementation leveraged \thesystem's module API.

Totaling 139 lines of code, the RollJam module implements the \lstinline{onUserCommand()} function to handle "start/stop" commands, and to allow the user to configure the number of packet repetition, and which radio to use for jamming and which one for listening.
The "start" command puts the listening radio in receive mode (\lstinline{setMode(RX, listenRadio)}) and calls the \lstinline{setMode(JAM, jamRadio)} on the jamming radio.
Any valid incoming packets will trigger the \lstinline{onPacketReceived(pkg)} function, implemented by the module.
If enough valid data is received, the jamming radio is idled and the listening radio is used to repeat the last valid rolling code (\lstinline{listRadio->transmit(pkt)}).

\subsection{RF Hacking Contests and Trainings}
\label{sec:cts}
Capture the flag (CTF) competitions have fully embraced RF hacking challenges, to the point that we can participate to RF-only contests, with Capture the Signal\footnote{\url{https://cts.ninja}} and Hack-a-Sat\footnote{\url{https://www.hackasat.com/}} being the most recent ones.

While reverse engineering offline signal captures is a useful learning experience, real-world targets will always be physical, interactive devices with real radio transceivers.
\thesystem makes it easy to provision interactive devices thanks to its scripting capabilities and firmware modularity, while maintaining fine-grained control on the RF protocol, which is an essential aspect for contests and trainings, because they require progressively more difficult challenges.

For example, we used \thesystem to build a remotely-programmable, battery-powered beaconing device that we hid in a conference room, activated it during a presentation, and asked the attendees to start inspecting the spectrum "around a certain frequency".
We asked them to locate the beacon, on which we physically printed a secret key, which they to transmit to the same device to unlock the next-level challenge.

Thanks to its cross-platform backend protocols (e.g., MQTT, including AWS IoT's MQTT), it is very easy to integrate \thesystem nodes into a web service, for example to emulate a complex IoT device with a web frontend.
\thesystem's firmware is already delivered via automated, "dockerized" builds, which make it suitable for automated, scalable deployments on multiple, distributed notes.

\section{Experimental Evaluation}
\label{sec:experiments}
While that the most useful aspect of \thesystem is its flexibility, the system must be robust in order to be usable.
Our first goal was to show that \thesystem can automatically capture sub-GHz transmissions in real time.
Then, we investigated the accuracy the automatic signal clamping feature, so that users know what to expect from it.

\subsection{End-to-end Experiment}
\label{sec:end-to-end}
We used a key fob to generate a reference signal, which we captured with a BladeRF SDR and decoded it with URH to know the baseline RF parameters.
Among the key fobs in \Cref{tab:key fobs}, we choose the Audi's HELLA FS12A70, because it transmits the shortest packet, which will put our auto-clamping module under time pressure\footnote{Video demo using the frequency and bitrate recovery on various key fobs: \url{https://youtu.be/36Bt8un_Y-Y}}.

We pressed the key fob button 50 times while the dongle was actively searching for valid transmissions.
We used a regular expression to parse the log and extracted the running time for both the frequency finder and the bitrate-estimation modules, and the recovered frequency and bitrate values.
We correctly identified and decoded 43 over 50 transmitted packets, which is a very positive result outside a Faraday cage.
The remaining 7 packets were wrongly decoded due to wrongly inferred parameters.

\begin{table}[b]
	\centering
  \resizebox{\columnwidth}{!}{%
	\begin{tabular}{|r||c|c||c|c|} 
		\hline
		                   & $t_{freq}$ [ms] & $t_{br}$ [ms] & $Freq$ [$MHz$] & $Br$ [$kbps$] \\
		\hline
		\hline
    Baseline           &                   &                 & \textbf{434.42}& \textbf{3.40}          \\ 
		\hline
		\hline
		Mean               & 22.55             & 10.26           & 434.48         & 3.80          \\ 
		\hline
		Mode               & 22.56             & 10.21           & 434.45         & 3.43          \\ 
		\hline
		First Quartile     & 22.53             & 10.19           & 434.42         & 3.38          \\ 
		\hline
		Third Quartile     & 22.57             & 10.35           & 434.46         & 3.45          \\
		\hline
		Standard Deviation & 0.08              & 0.13            & 0.10           & 1.52          \\  [1ex] 
		\hline
	\end{tabular}}
	\caption{Speed and accuracy on key fobs (\Cref{sec:end-to-end}).}
	\label{table:exp1}
\end{table}

\paragraph{Results} The results in \Cref{table:exp1} are aligned to our theoretical estimations.
The measured time for frequency estimation is about 1.5 ms greater than the expected.
We traced this back to the native radio driver performing some additional SPI transactions to handle special configurations.
For example, every time a packet is received, the driver queries the radio for the CRC configuration.
We optimized some of these transactions and we plan to review the rest of the native drivers for optimization.
The overall running time is $\leq 33$ ms, which is shorter than the preamble.

The mean values are affected by the 7 incorrect estimations.
We experimentally validated that a standard deviation up to 0.2 MHz is not posing issues to the decoding phase. Thus any estimated frequency between 434.2 and 434.6 MHz is considered valid.

The estimated mean bitrate is affected by outliers because this step runs after the frequency estimation.
For this reason, an incorrectly estimated frequency almost always implies a wrong bitrate value.
This explains the high standard deviation value despite the accurate mode.
Moreover, we experimentally validated that a bitrate deviation up to 0.3 kbps does not affect the decoded signal.

\subsection{Frequency Recovery Accuracy}
\label{sec:freq-tuning-accuracy}
We first focused on the frequency recovery part, using the second embedded radio to transmit the reference signal.
Although the transmitted payload is irrelevant for this experiment, we used the packet of one of our key fobs in order to have something realistic in terms of preamble length, sync-word, and total packet length.

To run the experiment we wrote a small script (see \Cref{sec:scripting-freq-tuning}) that tunes both radios at 432--437 Mhz, transmits with the first radio and collects any packets received by the second radio, along with the exact, detected carrier frequency.

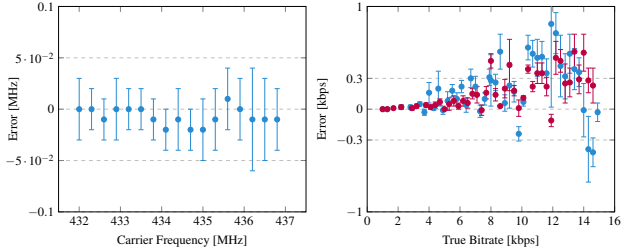
\begin{figure}[t]
	\pgfplotstableread{
		x         y     y-max    y-min
		432.0   -0.0    0.03    0.03
		432.3   0.0    0.02    0.02
		432.6   -0.01    0.02    0.02
		432.9   -0.0    0.03    0.03
		433.2   -0.0    0.02    0.02
		433.5   -0.0    0.02    0.02
		433.8   -0.01    0.02    0.02
		434.1   -0.02    0.02    0.02
		434.4   -0.01    0.03    0.03
		434.7   -0.02    0.02    0.02
		435.0   -0.02    0.03    0.03
		435.3   -0.01    0.03    0.03
		435.6   0.01    0.03    0.03
		435.9   0.0    0.03    0.03
		436.2   -0.01    0.05    0.05
		436.5   -0.01    0.04    0.04
		436.8   -0.01    0.03    0.03
    }{\secondexperimentb}
    
\pgfplotstableread{
	x         y     y-max    y-min
	1.0   0.0    0.01    0.01
	1.3   -0.0    0.01    0.01
	2.2   0.02    0.02    0.02
	2.8   0.02    0.02    0.02
	3.4   0.05    0.02    0.02
	3.7   -0.03    0.03    0.03
	4.0   0.16    0.1    0.1
	4.3   0.01    0.02    0.02
	4.6   0.2    0.17    0.17
	4.9   -0.01    0.04    0.04
	5.2   0.08    0.09    0.09
	5.5   0.18    0.06    0.06
	5.8   0.1    0.06    0.06
	6.1   0.22    0.06    0.06
	6.4   0.04    0.07    0.07
	6.7   0.3    0.09    0.09
	7.0   0.22    0.1    0.1
	7.3   -0.02    0.06    0.06
	7.6   0.1    0.06    0.06
	7.9   0.31    0.08    0.08
	8.0   0.28    0.25    0.25
	8.3   0.26    0.08    0.08
	8.6   0.56    0.17    0.17
	8.9   0.06    0.08    0.08
	9.2   0.23    0.22    0.22
	9.5   0.17    0.1    0.1
	9.8   -0.24    0.07    0.07
	10.1   0.07    0.04    0.04
	10.4   0.6    0.12    0.12
	10.7   0.54    0.13    0.13
	11.0   0.5    0.18    0.18
	11.3   0.51    0.15    0.15
	11.6   0.35    0.14    0.14
	11.9   0.83    0.81    0.81
	12.2   0.74    0.2    0.2
	12.5   0.42    0.3    0.3
	12.8   0.32    0.22    0.22
	13.1   0.54    0.19    0.19
	13.4   0.39    0.2    0.2
	13.7   0.36    0.14    0.14
	14.0   -0.01    0.26    0.26
	14.3   -0.39    0.32    0.32
	14.6   -0.42    0.14    0.14
	14.9   -0.03    0.09    0.09
  }{\thirdexperimentc}

  \pgfplotstableread{
    x         y     y-max    y-min
    1.0   0.0    0.01    0.01
    1.3   -0.0    0.01    0.01
    1.7   0.01    0.01    0.01
    2.2   0.02    0.02    0.02
    2.9   0.01    0.02    0.02
    3.2   0.03    0.02    0.02
    3.8   0.04    0.02    0.02
    4.1   0.02    0.02    0.02
    4.4   0.04    0.03    0.03
    4.7   0.07    0.03    0.03
    5.0   0.0    0.0    0.0
    5.3   0.05    0.06    0.06
    5.6   0.08    0.05    0.05
    5.9   0.03    0.03    0.03
    6.2   0.08    0.05    0.05
    6.5   0.06    0.05    0.05
    6.8   0.15    0.05    0.05
    7.1   0.14    0.08    0.08
    7.4   -0.01    0.05    0.05
    7.7   0.16    0.06    0.06
    8.0   0.47    0.07    0.07
    8.3   0.14    0.06    0.06
    8.6   0.03    0.0    0.0
    8.9   0.2    0.09    0.09
    9.2   0.43    0.25    0.25
    9.5   0.18    0.05    0.05
    9.8   0.01    0.08    0.08
    10.1   0.11    0.02    0.02
    10.4   0.39    0.04    0.04
    10.7   0.22    0.05    0.05
    11.0   0.35    0.1    0.1
    11.3   0.35    0.1    0.1
    11.6   0.22    0.09    0.09
    11.9   -0.11    0.06    0.06
    12.2   0.5    0.16    0.16
    12.5   0.47    0.11    0.11
    12.8   0.25    0.19    0.19
    13.1   0.26    0.14    0.14
    13.4   0.56    0.17    0.17
    13.7   0.29    0.14    0.14
    14.0   0.55    0.18    0.18
    14.3   0.28    0.12    0.12
    14.6   0.23    0.17    0.17
    }{\thirdexperimentd}
		  
	\begin{tikzpicture}[scale=0.48]
		\begin{axis}[
				xlabel={Carrier Frequency [MHz]},
				ylabel={Error [MHz]},
				xmin=431.5, xmax=437.5,
				ymin=-0.10, ymax=0.10,
				ytick={-0.10, -0.05, 0, 0.05, 0.10},
				xtick={432,433,434,435,436,437},
				legend pos=north west,
				ymajorgrids=true,
				grid style=dashed,
			]
			\addplot [
				color=blue,
				only marks,
				mark=*,
			] 
			plot [error bars/.cd, y dir=both, y explicit]
			table [y error plus=y-max, y error minus=y-min] {\secondexperimentb};
		\end{axis}
  \end{tikzpicture}
  \begin{tikzpicture}[scale=0.48]
    \begin{axis}[
        xlabel={True Bitrate [kbps]},
        ylabel={Error [kbps]},
        xmin=0, xmax=16,
        ymin=-1, ymax=1,
        ytick={-1,-0.3,0,0.3,1},
        xtick={0,2,4,6,8,10,12,14,16},
        legend pos=north west,
        ymajorgrids=true,
        grid style=dashed,
      ]
      \addplot [
        color=blue,
        only marks,
        mark=*,
      ] 
      plot [error bars/.cd, y dir=both, y explicit]
      table [y error plus=y-max, y error minus=y-min] {\thirdexperimentc};

      \addplot [
        color=purple,
        only marks,
        mark=*,
      ] 
      plot [error bars/.cd, y dir=both, y explicit]
      table [y error plus=y-max, y error minus=y-min] {\thirdexperimentd};
          
    \end{axis}
  \end{tikzpicture} 

  \caption{Carrier frequency peak detection and bitrate recovery accuracy (oversampling at 30kbps and 60kbps).}
  \label{fig:auto-tuning-accuracy}
\end{figure}

\paragraph{Results}
The results shown in \Cref{fig:auto-tuning-accuracy} (left) show that the algorithm is able to detect the frequency accurately, with an average error of \SI{0.02}{MHz}, and an average standard deviation of \SI{30}{kHz}.
This result is in line with our expectation since the narrowest receiver filter bandwidth for the CC1101 is \SI{58}{kHz}.
It follows that the error decreases while approaching the center of the receiver bandwidth, and increases up to \SI{30}{kHz} when moving away from it.
Note that a \SI{30}{kHz} error is acceptable: Recall that the RollJam attack works by jamming the receiver at \SI{50}{kHz} off the exact tuning frequency.

\subsection{Bitrate Recovery Accuracy}
\label{sec:bitrate-tuning-accuracy}
We used once again a script (see \Cref{sec:scripting-bitrate-tuning}) to automate the TX/RX of packets from \num{1} to \SI{15}{kbps}.
Following Nyquist–Shannon theorem, we oversampled at \num{30} and \SI{60}{kbps}, so obtaining the data plotted in \Cref{fig:auto-tuning-accuracy} (right).

\paragraph{Results}
We obtained good precision at low bitrates and a slightly higher precision overall at \SI{60}{kbps}, as expected.
We manually validated that the CC1101 can successfully decode signal with up to \SI{0.3}{kbps} estimation error.
Therefore, oversampling at \SI{60}{kbps} guarantees no data loss, up to \SI{15}{kbps}.

The reason why it almost seems that Nyquist-Shannon doesn't hold is twofold. First, from \Cref{eq:brate} recall that shorter sequences of consecutive 1s (or 0s) have greater influence on the estimated bitrate. 
The limit is exactly the Nyquist frequency, where the number of consecutive symbols tends to 1, determining wide oscillations of the estimated value.
The greater is the resolution, the less influence a single error has.

Secondly, after clamping to a signal, most transceivers (including the CC1110) internally estimate the clock from the incoming symbols, using a re-synchronization routine, which run continuously to eliminate small discrepancies between the chosen and real bitrate.
Thus, if oversampling bitrate is not an integer multiple of the incoming signal bitrate, the transceiver will automatically try to adjust the bitrate to compensate for unexpected symbols.
This feature, which cannot be disabled, influences our estimation.
We were not aware of such feature and we will search if it can be disabled in some transceivers.

\section{Limitations and Future Work}
\label{sec:discussion}
Like most community-driven projects, \thesystem also is meant to be a complete, final solution.
We focused on providing solid foundations and a flexible firmware for people to build upon.
There are, however, some limitations that we think should be addressed in the near future.

\paragraph{Scanned Bandwidth Limit} The first phase of the frequency recovery algorithm loops through all regions in the configured band.
It's still far more efficient than looping on the single frequencies, because it splits the spectrum in to a number of regions, thus reducing the number of iterations, but if the range is very wide, the running time may be incompatible with an online signal capture scenario.
This is a technical limitation, so the easy solution is to just use a transceiver with a wider filtering bandwidth.
Alternatively, the algorithm could be extended to shard the band across multiple transceivers.

\paragraph{Better Modulation Guessing}
The bitrate estimation algorithm assumes OOK modulation. 
To support other modulations such as 2- or 4-FSK, we foresee two solutions.
The frequency finder should look for the strongest RSSI on $N$ frequencies, to infer the number of occupied frequencies.
For example, a single occupied frequency may denote OOK, while 2 or 4 occupied frequencies indicate 2- or 4-FSK.

\begin{figure}[t]
  \centering
  \includegraphics[width=0.8\columnwidth]{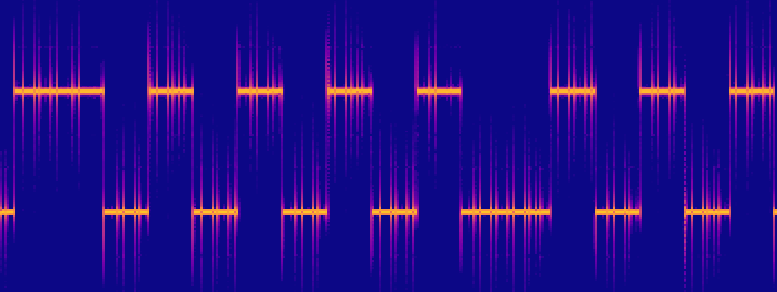}
  \caption{A single 2-FSK transmission, or two complementary OOK transmissions.}
  \label{fig:fsk}
\end{figure}

Alternatively, one can demodulate 2-FSK as if it was OOK, by using offset tuning.
FSK encodes symbols as discrete frequency changes, but this can be seen as two independent, complementary OOK transmissions on two distinct frequencies, as shown in \Cref{fig:fsk}.
Thus, once one of the two peak frequencies is found, the algorithm should tune "on the side" of just a few kHz and set the narrowest filter bandwidth in order to "see" only one of the two transmissions.
A second radio can be used to parallelize this task.

\section{Conclusions}
\label{sec:conclusions}
We presented the first truly modular RF system for security analysis of RF protocols.
\thesystem is midway between SDRs and RF dongles, by providing uniform high-level APIs to reconfigure an easy-to-customize hardware system, as well as the advanced features typically found in RF dongles such as the Yard Stick One.

We hope that it will encourage the community to implement new features, as we think that \thesystem can really change how we approach RF research and training.

\printbibliography

\clearpage

\appendix

\section{Screenshots}
\label{sec:screenshots}
\begin{figure}[h!]
  \centering
  \includegraphics[width=\columnwidth]{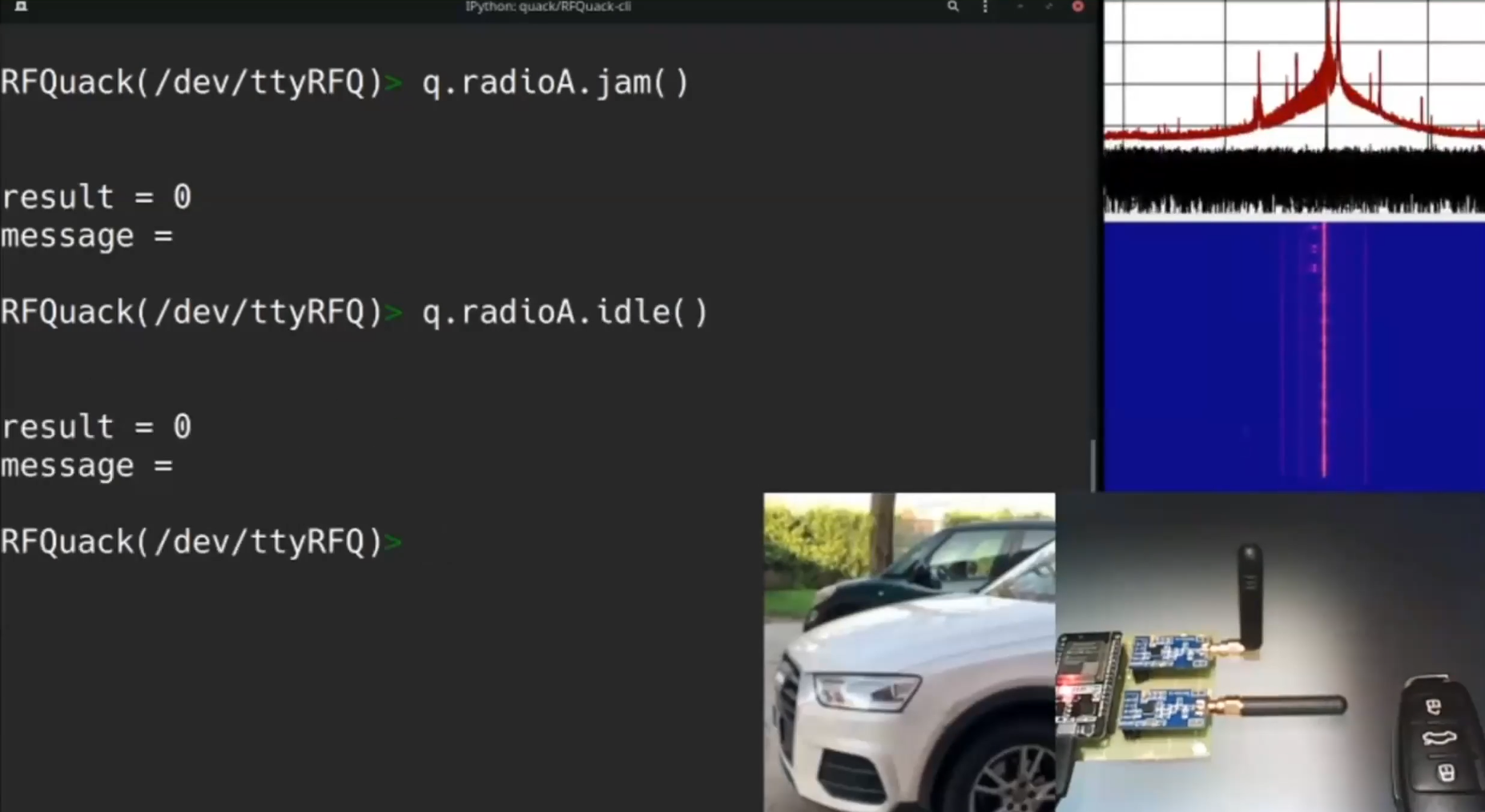}
  \caption{RollJam implementation on top of \thesystem in action, taken from durin presentation at CanSecWest.}
\end{figure}

\begin{figure}[h!]
  \centering
  \includegraphics[width=\columnwidth]{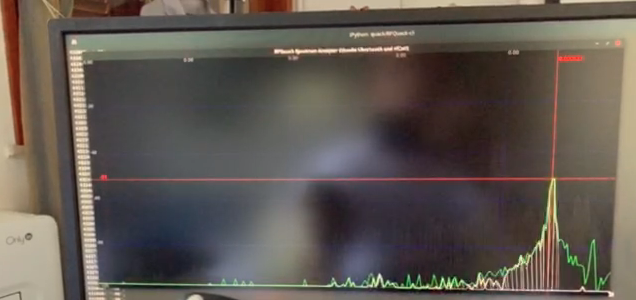}
  \caption{Spectrum slicing to find peaks and detect transmitting frequency, as described in \Cref{sec:peak-detection}.}
\end{figure}

\section{Transceiver-agnostic, Uniform API}
\label{sec:uniform-api}
\thesystem's support for multiple radios allows to use the functionalities of many different transceivers using the same API, which exposes the following groups of functions:

\paragraph{Radio Mode} Put the transceiver in standby, receiving, transmit, promiscuous, or jamming mode. In promiscuous mode, the radio will receive anything with RSSI above a user-configurable threshold (to avoid receiving pure noise), regardless of the preamble or sync-word values. The jamming mode just keeps transmitting, optionally leveraging transceiver-specific features to avoid busy loops (e.g., the CC11xx will keep transmitting as long as its TX FIFO is not empty).

\paragraph{RF Parameters} Set the carrier frequency, frequency deviation (for FSK modulations), RX bandwidth, and bitrate.

\paragraph{Packet Format} Set the fixed or variable packet length format, preamble length, sync-word value, and CRC check.

\paragraph{Carrier} Set the output power, get the last RSSI, determine if there is carrier at the tuned frequency.

\paragraph{TX and RX:} transmit a byte array or receive onto a byte array or queue. Optionally, activate RX loop and enqueue any received packet.

\paragraph{Register Access} Read or write values from or to the transceiver registers.

\section{Serialization and Transport Protocol}
\label{sec:serialization}
Protobuf is Google's cross-platform serialization protocol, which support most of the mainstream languages such as Python, Go, C, C++, Java.
It allows to create truly interoperable protocols and makes it easy to interface systems across very different languages and architectures.

To show an example of the type system of \thesystem, we describe the \lstinline{PacketModification} type, which is the core of \thesystem's packet-matching and modification engine.

\begin{lstlisting}[language={C},basicstyle=\ttfamily\scriptsize]
  // Tell the tool how to modify a byte of a packet.
  message PacketModification {
      // position in the packet
      optional uint32 position = 1;

      // pos = all indexOf(content)
      optional uint32 content = 2; 
  
      enum Op {
          // pkt[pos] = pkt[pos] & operand
          AND = 1;

          // pkt[pos] = pkt[pos] | operand
          OR = 2; 

          // pkt[pos] = pkt[pos] ^ operand
          XOR = 3;

          // pkt[pos] = ~pkt[pos]
          NOT = 4; 

          // pkt[pos] = pkt[pos] << operand
          SLEFT = 5; 

          // pkt[pos] = pkt[pos] >> operand
          SRIGHT = 6; 

          // pkt = payload + pkt
          PREPEND = 7; 

          // pkt = pkt + payload
          APPEND = 8; 

          // pkt = pkt[0 : pos] + payload + pkt[pos:pkt.size]
          INSERT = 9; 
      }
  
      optional Op operation = 3;
      optional uint32 operand = 4;
  
      // Apply only to packets matching a pattern
      optional string pattern = 5;
  
      // Bytes to append / prepend for OP = PREPEND | APPEND
      optional bytes payload = 6;
  }
\end{lstlisting}

The default \thesystem IPython shell has automatic autocompletion based on introspection of the Protobuf types.
This means that new Protobuf types are automatically "added" and supported by the console, without changing any code.

To achieve RPC functionalities, the Protobuf messages are prepended by a URI that identifies the command (similarly to MQTT topics) exchanged between the host and the dongle.
The resulting binary messages are transported over a serial connection (USB) and over MQTT for WiFi (and cellular).

\section{Console Scripting}
\label{sec:scripting}
In this section we provide some examples of \thesystem's IPython shell scripting capabilities, which we used in the case studies and experiments described in \Cref{sec:cases} and \ref{sec:experiments}.

\subsection{Frequency Recovery Experiment}
\label{sec:scripting-freq-tuning}
This script, mentioned in \Cref{sec:freq-tuning-accuracy}, tunes both radios at 432--437 Mhz, transmits with the first radio and collects any packets received by the second radio, along with the detected carrier frequency.

\begin{lstlisting}[language={Python},basicstyle=\ttfamily\scriptsize]
  q.radioA.rx()  # set first radio in RX mode
  q.guessing.start_freq = 432
  q.guessing.end_freq = 437
  q.guessing.start()
  values = dict()   
  for i in range(4320, 4373, 3):  
     freq = i/10 
     values[freq] = list() 
     q.radioB.set_modem_config(carrierFreq=freq)
     q.radioB.tx()  # second radio in TX mode
     time.sleep(1) 
     for times in range(0, 50): 
         q.data = []  # Clear recv packets buffer
         q.radioB.send(data=bytes.fromhex("aaa[....]666"))
         time.sleep(0.5)  # Wait for decoding
         values[freq].append(q.data[0].carrierFreq)  
\end{lstlisting}

\subsection{Bitrate Recovery Experiment}
\label{sec:scripting-bitrate-tuning}
This script, mentioned in \Cref{sec:bitrate-tuning-accuracy}, sets one radio in RX mode and enables the bitrate recovery loop.
Then it transmits the same packet at varying bitrate values, for 50 times each, letting the first radio clamp on the signal and estimate the bitrate.

\begin{lstlisting}[language={Python},basicstyle=\ttfamily\scriptsize]
 q.radioA.rx()  # set first radio in RX mode
 q.guessing.sampling_bitrate = 30  #Then, 60
 q.guessing.start()
 values = dict()

 for i in range(10, 150, 3):
     br = i/10 
     values[br] = list() 
     q.radioB.set_modem_config(bitRate=br) 
     q.radioB.tx() 
     q.sleep(1)
     for times in range(0, 50):     
         q.data = []  # Clear received packets
         q.radioB.send(data=bytes.fromhex("aaa[....]666"))
         time.sleep(0.5)  # Wait for decoding
         # Decoded packets get stored in q.data
         values[br].append(q.data[0].bitRate)
\end{lstlisting}

\subsection{Isolate 2.4GHz Valid Frames}
\label{sec:24ghz-analysis}
This script set the nRF24 2.4GHz frontend in promiscuous mode at \SI{2000}{kbps}, collects all the sync words and count their occurrences.
Using the most frequent sync words, we queried the FCC database and found a lead that helped us isolate the traffic of the target device.

\begin{lstlisting}[language={Python},basicstyle=\ttfamily\scriptsize]
  q.radioA.set_modem_config(bitRate=2000, isPromiscuous=True)
  q.radioA.set_packet_len(isFixedPacketLen=True, packetLen=32)  # max
  q.radioA.rx()

  # capture anything within the range
  for freq in range(2405, 2474):
    q.radioA.set_modem_config(carrierFreq=freq)

  # isolate all sync word and rank
  sw = list(map(lambda x: x.data.hex()[0:10], q.data))
  counter = Counter(sw)
\end{lstlisting}

\end{document}